\documentclass[a4paper]{article}
\usepackage{amssymb}
\usepackage{graphicx}
\usepackage{url}
\newcommand{\keywords}[1]{\par\addvspace\baselineskip
\noindent\keywordname\enspace\ignorespaces#1}
\PassOptionsToPackage{svgnames}{xcolor}
\usepackage[utf8]{inputenc} % NEEDED for UTF-8 characters such as accented.

\usepackage{libertine} % NEEDED for textsf and textsc at the same time (not all sf fonts have small capitals)
\def\textsf#1{{\biolinum #1}}
\def\scsf#1{{\scshape \biolinum #1}} % THIS line does the sf and sc
\usepackage{lmodern} % NEEDED if lmodern is to be the main font (otherwise, libertine is used).
\usepackage[scaled=0.95]{inconsolata} % NOT needed but looks better for texttt

\usepackage[titlenumbered%
,ruled%
]{algorithm2e} % NEED that for my algorithms

\usepackage[%
framemethod=%
TikZ
,nobreak=false
]{mdframed} % NEED that for my listings
\usepackage{caption,listings} % NEED that for my listings

\usepackage{wrapfig} % NEED that to wrap one figure in text.

\usepackage{cite} % NEED that to order citation automatically, can be removed though, but citation order is meant to be increasing.

\usepackage{booktabs} % NEED that for my tables

\usetikzlibrary{shapes,matrix,arrows}

\mdfsetup{roundcorner=5%
,backgroundcolor=LightBlue!2!white%
,linecolor=LightBlue!60!black%
,leftmargin=0pt,rightmargin=0pt
,innerleftmargin=8pt,innerrightmargin=5pt
,skipabove=0.9\topskip
,skipbelow=-0.9\topskip
}

\BeforeBeginEnvironment{lstlisting}%
{\begin{mdframed}[nobreak=false]
	\vskip-0.6\baselineskip
}
\AfterEndEnvironment{lstlisting}{\vspace{-0.15\baselineskip}\end{mdframed}\vspace{1.0\baselineskip}\par\nobreak}
\def\ifempty#1{\def\temparg{#1}\ifx\temparg\empty}
\newcommand{\lstcapt}[2][]{%
\vspace{-0.5\baselineskip}%
\ifempty{#1}%
\captionof{lstlisting}{#2}%
\else%
\captionof{lstlisting}[#1]{#2}%
\fi%
\vspace{0.3\baselineskip}%
}

\def\num#1{#1}

\def\ldots{...}
\newcommand\Zb{\mathbf{Z}}
\newcommand\modring[1]{\ensuremath\Zb{/ #1\Zb}}
\newcommand\gnuplot{\textsf{gnuplot}\xspace}
\newcommand\linbox{\textsf{LinBox}\xspace}
\newcommand\mul{\texttt{mul}\xspace}
\newcommand\apply{\texttt{apply}\xspace}
\newcommand\spmv{\textrm{SpMV}\xspace}
\newcommand\cpp{\texttt{C++}\xspace}

\newcommand\fgemm{\texttt{fgemm}\xspace}
\newcommand\fflas{\scsf{Fflas}\xspace}
\newcommand\ffpack{\scsf{Ffpack}\xspace}
\newcommand\fflasffpack{\scsf{Fflas--Ffpack}\xspace}
\newcommand\givaro{\textsf{Givaro}\xspace}
\newcommand\applin{\textsf{BlackBox}\xspace}
\newcommand{\cf}{\mbox{\emph{cf.}}\xspace}
\newcommand{\eg}{\mbox{\emph{e.g.}}\xspace}
\newcommand{\ie}{\mbox{\emph{i.e.}}\xspace}
\newcommand{\etc}{\mbox{\emph{etc.}}\xspace}
\newcommand{\vs}{\mbox{\emph{vs.}}\xspace}
\newcommand{\adhoc}{\mbox{\emph{ad hoc.}}\xspace}

\newcommand{\blas}{\scsf{Blas}\xspace}
\newcommand\td{\textsuperscript{\textdagger}}
\newcommand\tdd{\textsuperscript{\textdaggerdbl}}

{%\theoremstyle{remark}
}
{
}

\usepackage[
	breaklinks%
	,colorlinks%
	,linkcolor=MidnightBlue%
	,anchorcolor=cyan%
	,urlcolor=DarkGreen
	,citecolor=blue%
	,bookmarks=false%
]{hyperref}
\usepackage[capitalize]{cleveref} % NEED that for easy referencing environemnts.

\begin{document}

% \mainmatter

%===== DO NOT MODIFY BEGIN============================
% \documentclass[runningheads,a4paper]{llncs}
% \usepackage{amssymb}
% \setcounter{tocdepth}{3}
% \usepackage{graphicx}
% \usepackage{url}
% \newcommand{\keywords}[1]{\par\addvspace\baselineskip
% \noindent\keywordname\enspace\ignorespaces#1}
%===== DO NOT MODIFY END==============================
% \input{Boyer_Brice/top}
%===== DO NOT MODIFY BEGIN============================
% \begin{document}

% \mainmatter
%===== DO NOT MODIFY END==============================

%%%%%%%%%%%%%%%%%%%%%%%%%%%%%%%
%%% Title
%%%%%%%%%%%%%%%%%%%%%%%%%%%%%%%

\title{Elements of Design for Containers and Solutions in the \linbox library \\
% }
% \subtitle{
	\large{Extended abstract}}
% \titlerunning{Design for Containers and Solutions in the \linbox library}
%
\author{Brice Boyer\\%
	\renewcommand{\thefootnote}{\fnsymbol{footnote}}
	\normalsize{Department of Mathematics, North Carolina State University,
	USA}\footnotemark[4]\\
	\email{bbboyer@ncsu.edu} \\[0.2em]
	Jean-Guillaume Dumas\\%
	\normalsize{Laboratoire J. Kuntzmann, Universit\'e de Grenoble,
	% 51, rue des Math\'ematiques, umr CNRS 5224, bp 53X, F38041 Grenoble,
	France}\footnotemark[5]\\
	\email{Jean-Guillaume.Dumas@imag.fr} \\[0.2em]
Pascal Giorgi \\%
\normalsize{LIRMM%
	, CNRS, Universit\'e Montpellier 2,
	% 161 rue ADA, F-34095 Montpellier,
France}
\footnotemark[5]\\
	\email{pascal.giorgi@lirmm.fr} \\[0.2em]
Cl\'ement Pernet \\%
	\normalsize{Laboratoire LIG, Universit\'e de Grenoble et INRIA,
	France}\footnotemark[5]\\
	\email{clement.pernet@imag.fr}\\[0.2em]
B. David Saunders \\%
	\normalsize{University of Delaware, Computer and Information Science Department,
	% Newark / DE / 19716,
	USA}%
	\footnotemark[4]\\
	\email{saunders@udel.edu}%
}
% \authorrunning{Boyer--Dumas--Giorgi--Pernet--Saunders}%
% \authorrunning{B. Boyer, J-G. Dumas, P. Giorgi, C. Pernet, B. D. Saunders}
\def\email#1{\href{mailto:#1}{#1}}
\date{}
\newif\ifAbstractOnly
% \AbstractOnlytrue
\AbstractOnlyfalse
\maketitle
\def\keywords#1{\par\textbf{Keywords: }#1}
\footnotetext[4]{\footnotesize This material is based on work supported in part by
	the National Science Foundation under Grant CCF-1115772 (Kaltofen) and Grant CCF-1018063 (Saunders)}
\footnotetext[5]{\footnotesize This material is based on work supported in part by the
          Agence Nationale pour la Recherche under Grant
          ANR-11-BS02-013 HPAC (Dumas, Giorgi, Pernet).}
\begin{abstract}
	We describe in this paper new design techniques used in the \cpp exact
	linear algebra library \linbox, intended to make the library
	safer and easier to use, while keeping it generic and efficient.
	%
	% \par
	% \\
	%
	First, we review the new simplified structure for containers, based on
	our \emph{founding scope allocation} model. We explain design choices
	and their impact on coding: unification of our matrix classes, clearer
	model for matrices and submatrices, \etc
	%
	% \par
	% \\
	%
	Then we present a variation of the \emph{strategy} design pattern that
	is comprised of a controller--plugin system: the controller (solution)
	chooses among plug-ins (algorithms) that always call back
	the controllers for subtasks. We give examples using the solution \mul.
	%
	% \par
	% \\
	%
	Finally we present a benchmark architecture that serves two purposes:
	Providing the user with easier ways to produce graphs;
	Creating a framework for automatically tuning the library and supporting
	regression testing.
\end{abstract}
\ifAbstractOnly
% \ifdefined\doabstract
\else
%%%%%%%%%%%%%%%%%%%%%%%%%%%%%%%
% Intro
%%%%%%%%%%%%%%%%%%%%%%%%%%%%%%%
%
% \input{Boyer_Brice/1-intro}
\section{Introduction}
This article follows several papers and memoirs concerning \linbox%
\footnote{See \url{http://www.linalg.org}.}
(\cf{} \cite{Giorgi:2004:these,Turner:2002:these,Boyer:2012:these,Dumas:2002:icms,Dumas:2010:lbpar})
and builds upon them.
\linbox is a \cpp template library for fast and exact linear algebra, designed with generality
and efficiency in mind.
% The main concepts include:
% \begin{itemize}[---]
	% \item A hierarchy fields--containers--algorithms--solutions
	% \item Matrix containers are Blackboxes
	% \item RAII reentrant style
% \end{itemize}
%
% \par
%
The \linbox library is under constant evolution, driven by new problems and
algorithms, by new computing paradigms, new compilers and architectures. This
poses many challenges: we are incrementally
updating the \emph{design} of the library towards a \textsf{2.0} release.
The evolution is also motivated by developing a high-performance mathematical
library available for researchers and engineers that is easy to use and help
produce quality reliable results and quality research papers.
\par
Let us start from a basic consideration: we show in the \Cref{tab:sloc} the
increase in the ``lines of code'' size\footnote{Using \textsf{sloccount}, available at
\url{http://sourceforge.net/projects/sloccount/}.}
of \linbox and its coevolved dependencies \givaro and
\fflasffpack\footnote{symbol \td when \givaro is included and \tdd when
contains \fflasffpack}.
%
% \par
%
% \input{Boyer_Brice/tab_sloc}
\begin{table}[htbp]
	\begin{center}
	% \vspace{-1.1em}
	\small
		\renewcommand{\arraystretch}{0.82}
	\begin{tabular}{lcccccccccc}
		\toprule
		\linbox & \sf 1.0.0\td \tdd & \sf 1.1.0\td \tdd& \sf 1.1.6\tdd & \sf 1.1.7\tdd & \sf 1.2.0 & \sf 1.2.2 & \sf 1.3.0 & \sf 1.4.0\\
		loc ($\times \num{1000}$) & {\num{77.3}}& {\num{85.8}} & {\num{93.5}} & {\num{103}} & {\num{108}}  & {\num{109}} &        {\num{112}} & \num{135} \\
		\cmidrule(lr){1-9}
		\fflasffpack &n/a&n/a& n/a & \sf 1.3.3 & \sf 1.4.0 & \sf 1.4.3 & \sf 1.5.0 & \sf 1.8.0 \\
		loc & --- & ---& --- &\num{11.6} & {\num{23.9}} & {\num{25.2}} & {\num{25.5}} & \num{32.1}\\
		\cmidrule(lr){1-9}
		\givaro        & n/a& n/a & \sf 3.2.16        & \sf 3.3.3         &  \sf 3.4.3         & \sf 3.5.0         & \sf 3.6.0 & \sf 3.8.0 \\
		loc&---&---& \num{30.8} & {\num{33.6}}   & {\num{39.4}} & {\num{41.1}} & {\num{41.4}} &  \num{42.8} \\
		\midrule
		total  &  \num{77.3} & \num{85.8} & \num{124} & \num{137} & \num{171} & \num{175} & \num{179} & \num{210} \\
		\bottomrule \\
	\end{tabular}
	% \vspace{-0.5em}
	\caption{Evolution of the number of lines of code  in
		\linbox.
		% , \fflasffpack and \givaro (\td contains \givaro, \tdd
	% contains \fflasffpack).
	}
	\label{tab:sloc}
\end{center}
	% \vspace{-3.5em}
\end{table}%
This increase affects the library in several ways.  First, it demands a
stricter development model, and we are going to list some techniques we used.
%% SPLITTING
For instance, we have transformed \fflasffpack %
%
% \footnote{See \url{http://www.linalg.org/projects/fflas-ffpack/}.}
	%
(\cf{} \cite{Dumas:2008:Flas}) into a new standalone header library, resulting
in more visibility for the \fflasffpack project
% (\danger Singular ?)
and also in better structure and maintainability of the library.
% focusing the development areas more precisely.
%
%% FIXING
%
A larger template library is harder to manage. There is more difficulty
to trace, debug, and write new code. Techniques employed
for easier development include reducing
compile times, enforcing stricter warnings and checks, supporting more
compilers and architectures, simplifying and automating version number
changes, automating memory leak checks, and setting up buildbots to check the code
frequently.
% This demand on the developer
% side is also driven by the fact that code introduced by various one-timers
% needs to be maintained.
%
\par
This size increase also requires more efforts to make the library user friendly. For
instance, we have:
Developed %an \texttt{auto-install.sh}
scripts that install automatically the
latest stable/development versions of the trio, resolving version
dependencies;
Eased the discovery of \blas/\scsf{Lapack} libraries;
Simplified and sped up the checking process, covering more of the library;
% (\danger dave a word on make fullcheck or on the future feature matrix ---what
% solution, what field, what implementation we recommend/provide if at all ?);
%
Updated the documentation and distinguished user and developer oriented docs;
Added comprehensive benchmarking tools.
\par
Developing generic high performance libraries is difficult. We can find a
large literature on coding standards and software design references in (\cf{}
\cite{alexandrescu:01:modern,gamma:95:design,sutter:05:cpp,stroustrup1994design,Douglas:05:GPHP}),
and draw from many internet sources and experience acquired by/from free
software projects.
%
% \par
%
We describe advances in the design of \linbox in the next
three sections. We will first describe the new \emph{container} framework in
\Cref{sec:container}, then,
in \Cref{sec:matmul},
the improved \emph{matrix multiplication} algorithms
made by contributing special purpose matrix multiplication plugins, and, finally, we present the new \emph{benchmark/optimization}
architecture (\Cref{sec:bench}).
\par
% (\danger develop this § more later when I have more material.)
%
%%%%%%%%%%%%%%%%%%%%%%%%%%%%%%%
% Containers
%%%%%%%%%%%%%%%%%%%%%%%%%%%%%%%
%
%
% \input{Boyer_Brice/2-containers}
\section{Containers architecture}\label{sec:container}
\linbox is mainly conceived around the RAII (Resource Acquisition Is
Initialization, see \cite{stroustrup1994design}) concept with reentrant
function. We also follow the {founding scope allocation} model (or \emph{mother
model}) of \cite{Dumas:2010:lbpar} which ensures that the memory used by
objects is allocated in the constructor and freed only at its destruction. The
management of the memory allocated by an object is exclusively reserved to it.
\par
\linbox uses a variety of container types (representations) for matrix and vectors over fields and rings.
The fragmentation of the containers into various matrix and
blackbox types has been addressed and simplified. The many different matrix and
vector types with different interfaces has been reduced into only two
%(possibly essentially one in the future)
containers: \texttt{Matrix} and
\texttt{Vector}.
\subsection{General Interface for %Vectors and
Matrices}
First, in order to allow operations on its elements, a container is
parameterized by a field object (\Cref{code:clmat}), not the field's element
type. This is simpler and more general.
Indeed, the field element type can be inferred from a
\verb!value_type! type definition within the field type.
Then, the storage type is given by a second
template parameter that can use defaults,
\eg dense \blas matrices (stride
and leading dimension or increment), or some sparse format.
% what else can it be? It can default to other dense formats such as vectors of
% rows or anything.
%
% \input{Boyer_Brice/lst_container}
{
	\small
\begin{lstlisting}
template< class _Field, class _Storage = denseDefault >
class Vector ;
\end{lstlisting}
\lstcapt{Matrix or Vector classes in \linbox.\label{code:clmat}}
}
% template< class _Field, class _Storage = denseDefault >
% class Matrix ;
%
In the founding scope allocation model, we must distinguish containers that own (responsible for
dynamically allocated memory) and containers that share memory of another.
\texttt{SubMatrix} and \texttt{SubVector} types share the memory;
\texttt{Matrix} and \texttt{Vector} own it.
All matrix containers share the common \applin  interface
described in the next paragraphs, it accommodates both owner and sharer
container types, and defines the minimal methods required for a
template \applin matrix type:
% \Cref{code:bb}.
%
% \input{Boyer_Brice/lst_applin}
%
% \begin{description}
%
	% \item[Input/Output.]
\def\monitem#1{\par\textit{#1}\ }
\monitem{Input/Output.}
		Our matrix containers all read and write from Matrix Market
		format\footnote{See \url{http://math.nist.gov/MatrixMarket/}.}
		which is well established in the numerical linear algebra
		community and facilitates sharing matrices with other software
		tools.  The MatrixMarket header comment provides space for
		metadata about the provenance of a matrix and our interest in
		it.  However, because of our many entry domains and matrix
		representations, extensions are necessary to the MatrixMarket
		format.  For instance, the header comment records the modulus
		and irreducible polynomial defining the representation of a
		matrix over GF($p^e$).
% we haven't committed to this as a MM variant, should we? bb: yes!..
		We can further adapt the header to suit our needs, for instance
		create new file formats that save space (\eg CSR fashion saves
		roughly a third space over COO, \cf Harwell-Boeing format).
		Structured matrices (Toeplitz, Vandermonde, \etc) can have file
		representations specified.
%
%
	% \item[Apply method.]
		\monitem{Apply method.}
		This is essential in the \applin interface (\Cref{ssec:apply,sec:matmul}).
%
	% \item[Rebind/Conversions.]
		\monitem{Rebind/Conversions.}
		In addition to the rebind mechanism (convert from one field to the other), we add
	 conversion mechanisms between formats, for instance all
	 sparse matrix formats can convert to/from CSR format:  this `star'
	 mechanism can simplify the code (to the expense of memory usage) and
	 may speed it up when some central formats  are
	 well tuned for some task.
	 % Such tasks could be
	 % read/write from a file for instance, or update the matrix.
%
% \end{description}
	 % \paragraph{}
	 \par
This is a common minimal interface to all our matrix containers that can be
used by all algorithms.
% Additional elements to a container can be added, and flagged
% with a trait.
This interface provides the basic {\em external}
functionality of a matrix as a
 ``linear mapping'' (black box).
This interface is shared by: \emph{dense}
containers (\blas-like,\ldots); \emph{permutation} containers (compressed
\scsf{Lapack} or cycle representation); \emph{sparse} containers (based on
common formats or on STL containers such as {\tt map}, {\tt deque},\ldots);
\emph{structured} containers ({\tt Diagonal}, {\tt Hankel}, {\tt Butterfly},\ldots);
\emph{compound} containers ({\tt Compose}, {\tt Submatrix},\ldots).
Additional functions of a container can be added, and flagged with a trait, for
example those that support internal changes as for Gaussian elimination.
\subsection{The \texttt{apply} method}\label{ssec:apply}
%
% We will discuss the \texttt{apply} function in \Cref{sec:apply}.
% Both vec/mat
%
\par
The \texttt{apply} method (left or right) is arguably the most important
feature in the matrix interface and the \linbox library. It performs what a
linear application is defined for: apply to a vector (and by extension  a block
of vectors, \ie a matrix).
\par
We propose the new interface (\Cref{code:apply}), where {\tt \_In} and {\tt
\_Out} are vector or matrices, and {\tt Side} is {\tt Tag::Right} or
{\tt Tag::Left}, whether the operation $y \gets A^{\top} x$ or  $y \gets A x$ is
performed. We also generalize to the operation $y \gets \alpha A x + \beta y$.
% as it is often handy and efficient to allow for accumulation.
%
% \input{Boyer_Brice/lst_apply}
{
\small
% // y = A.x
\begin{lstlisting}
template< class _In, class _Out  >
_Out& apply(_Out &y, const _In& x, enum Side) ;
\end{lstlisting}
\lstcapt{Apply methods.\label{code:apply}}
}
% // y = alpha y + beta A.x
% template< class _In, class _Out  >
% _Out& applyAcc(_Out &y, const Element& alpha, const _In& x, const Element& beta, enum Side) ;
%
This method is fundamental as it is the building block of the \applin
algorithms (for instance block-Wiedemann) and as the matrix
multiplication, main operation in linear algebra, needs to be extremely efficient (\Cref{sec:matmul}).
%
%
% \par
%
The implementation of the apply method can be left to a \mul solution,
which can include a helper/method argument if the apply
parameters are specialized enough.
%
% \Cref{code:applymul}.
%
% Usually,  the {\tt apply} method for
% containers is strongly dependent on the representation of the matrix. In
% the case of Hybrid containers, it will simply be calls to the {\tt apply}
% method of the children.
%
%
%
%%%%%%%%%%%%%%%%%%%%%%%%%%%%%%%
% Mat Mul
%%%%%%%%%%%%%%%%%%%%%%%%%%%%%%%
%
%
% \input{Boyer_Brice/3-matmul}
\section{Improving \linbox matrix multiplication}\label{sec:matmul}
We propose a \emph{design pattern} (the closest pattern to our knowledge is the
\emph{strategy} one, see \cite[Fig 2.]{Cung:2006:TC}) in \Cref{ssec:plugin} and
we show a variety of new algorithms where it is used in the \mul  solution
(\Cref{ssec:algmul}).
\subsection{Plugin structure}\label{ssec:plugin}
We propose in \Cref{fig:diag:patt} a generalization of the \emph{strategy
  design pattern} of \cite[Fig 2.]{Cung:2006:TC}, where distinct
algorithms (modules) can solve the
same problem and are combined, recursively, by a controller.
The main advantage of our pattern is that the modules always call
the controller of a function so that the best version will be chosen
at each level.
% Besides modules can be easily added as \emph{plugins}.
An analogy can be drawn with dynamic systems --- once the controller sends a
correction to the system, it receives back a new measure that allows for a new
correction.
\par
\begin{wrapfigure}{r}{0.41\textwidth}
% \begin{figure}[htbp]
	% \vspace{-2.2em}
	\begin{center}
	\small
\tikzstyle{block} = [draw, fill=blue!20, rectangle,
    minimum height=1.7em, minimum width=5.5em, rounded corners]
\tikzstyle{input} = [coordinate]
\tikzstyle{output} = [coordinate]
\tikzstyle{fleche} = [draw,->,shorten <=3pt, shorten >=3pt,thick]
% \tikzstyle{pinstyle} = [pin edge={to-,thin,black}]
	% \centering
\begin{tikzpicture}[auto, node distance=2.0cm,>=latex']
        \node [input, name=input] (input) {};
        \node [block, right of=input] (controller) {Controllers};
        \node [output, right of=controller] (output) {};
        \node [block, below of=controller] (modules) {Modules};
        \draw [fleche] (input) to [near start] node {input} (controller);
        \draw [fleche] (controller) to [near end] node {output} (output);
        \draw [fleche] (controller.300) to [bend left,near end] node [right] {call} (modules.60);
        \draw [fleche] (modules.120) to [bend left,near end] node [left] {call back} (controller.240);
\end{tikzpicture}
	% \vspace{-1em}
\caption{Controller/Module design pattern}
\label{fig:diag:patt}
\end{center}
	% \vspace{-2.2em}
% \end{figure}
\end{wrapfigure}
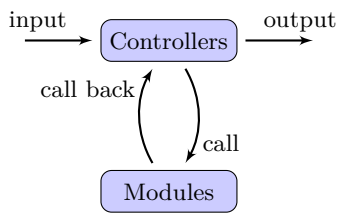
For instance, we can write (\Cref{fig:seuil}) the standard cascade algorithms
(see \cite{Dumas:2008:Flas}) in that model. Cascade algorithms are used to combine
several algorithms that are switched using thresholds, ensuring better
efficiency than that of any of the algorithms individually.
%
% \par
%
%
This method allows for the reuse of modules and ensures efficiency.
It is then possible to adapt to the architecture, the available modules,
the resources. The only limitation is that the choice of module
must be fast.
%
% \par
%
On top of this design, we have Method objects that allow caller selection
of preferred algorithms, shortcutting the strategy selection.
\begin{figure}[h]
	% \vspace{-1em}
	\small
        \hfil
	\begin{minipage}[t]{0.45\textwidth}
                \begin{algorithm}[H]
                        \caption{\texttt{Algo}: controller}
                        \label{alg:controle}
                        \KwIn{$A$ and $B$, dense, with resp. dimensions $n\times
                        k$ and $k\times n$.}
			\KwIn{$H$ Helper}
                        \KwOut{$C = A\times B$}
			\eIf{$\mathrm{min}(m,k,n)<H.{\tt threshold}()$}{
				% {\tt BaseCase} K() ; \\
				{\tt Algo}(C,A,B,{\tt BaseCase}()) ; %\tcc{\small{\color{gray}fast BLAS}}
			}
			{
				% {\tt RecursiveCase} H() ; \\
				{\tt Algo}(C,A,B,{\tt RecursiveCase}())
			}
			% \KwRet C \;
                \end{algorithm}
        \end{minipage}
        \hfil
	\begin{minipage}[t]{0.49\textwidth}
                \begin{algorithm}[H]
                        \DontPrintSemicolon
                        \caption{\texttt{Algo}: recursive module}
                        \label{alg:action}
                        \KwIn{$A$, $B$, $C$ as in controller.}
			\KwIn{$H$, {\tt RecursiveCase} Helper}
                        \KwOut{$C = A\times B$}
                        Cut $A$,$B$,$C$ in $S_i, T_i$\;
                        \ldots \;
                        $P_i = \mathtt{Algo}(S_i,T_i,H)$ \;
                        \ldots \;
			% \KwRet C \;
                \end{algorithm}
        \end{minipage}
        \hfil
        \caption{Conception of a recursive controlled algorithm}
        \label{fig:seuil}
	% \vspace{-2em}
\end{figure}
% \par
%
% \danger timing old fgemm/plugin fgemm with no noticeable change ?
%
\par
This infrastructure supports modular code. For instance,
\fflasffpack has seen major  modularization (addition, scaling,
reduction,\ldots) Not only does it enable code to be hardly longer than
the corresponding pseudocode listings, \cite{Boyer:2009:sched}, (compared to
$\approx 2.5\times$ on some routines before) but it also automatically brings
performance, because we can separately improve individual modules and
immediately have the benefit throughout the whole library.
%% following could be in, but perhaps cut now for reducing to 8 pages
%Also, this reduces the lines of code, hence the
%probability for bugs, eases their tracing/tracking, and allows for more
%unit tests. Modularizing the code comes at almost no cost because we may add
%$O(1)$  operations that: Do not cost much compared to $O(n^2)$ or more
%complexity of the modules; Allow early decisions and terminations (\eg by
%testing against $0$, $\pm 1$, checking leading dimensions or increments);
%Allow better code (AVX, SSE, copy--cache friendly operation--copy back,
%representation switching,\ldots)
%
\subsection{New algorithms for the \mul solution}\label{ssec:algmul}
New algorithms and techniques improve on matrix multiplication
in several  ways: reducing memory consumption, reducing runtime,
using graphics capabilities, generalizing the \blas to integer
routines.
\def\monitem#1{\par \textit{#1}\ }
\monitem{Reduced memory.}
The routine \fgemm in \fflas uses by default the classic schedules for the multiplication
and the product with accumulation (\cf \cite{Boyer:2009:sched}), but we also
implement the low memory routines therein. The new algorithms are competitive
and can reach sizes that were limiting.
%
% \par
%
One difficulty consists in using the memory contained in a submatrix of
the original matrix, that one cannot free or reallocate.
\monitem{Using Bini's approximate formula.}
In \cite{BD:2014:Bini}, we use Bini's approximate matrix multiplication formula
to derive a new algorithms that is more efficient that the Strassen--Winograd
implementation in \fgemm by $\approx 5-10\%$ on sizes \num{1500}--\num{3000}.
This is a cascade of Bini's algorithm and Strassen--Winograd algorithm and/or
the naïve algorithm (using \blas). The idea is to analyze precisely the error
term in the approximate formula and make it vanish.
\monitem{Integer \blas.}
In order to provide fast matrix multiplication with multiprecision integers, we
rely on multimodular approach through the Chinese remainder theorem. Our
approach is to reduce as much as possible to \fgemm. Despite, the existence of
fast multimodular reduction (resp.\ reconstruction) algorithm
\cite{VonzurGathen:1999:MCA}, the naïve quadratic approach can be reduced to
\fgemm which makes it more efficient into practice.  Note that providing
optimized fast multimodular reduction remains challenging. This code is
directly integrated  into \fflas.
\monitem{Polynomial Matrix Multiplication over small prime fields.}
The situation is similar to integer matrices since one can use
evaluation/interpolation techniques through DFT transforms. However, the
optimized Fast Fourier Transform of \cite{Harvey:2014}  makes fast evaluation
(resp.\ interpolation) competitive into practice. We thus rely on this scheme
together with \fgemm for pointwise matrix multiplications. One can find some
benchmark of our code in \cite{GioLeb14}.
\monitem{Sparse Matrix--Vector Multiplication.}
For sparse matrices a main issue is that the notion of \emph{sparsity}
is too general \vs the specificity of real world sparse matrices: the
algorithms have to adapt to the shape of the sparse matrices.
There is a huge literature from numerical linear algebra  on \spmv (Sparse
Matrix Vector multiplication) and on sparse matrix formats, some of which are
becoming standard (COO, CSR, BCSR, SKY,\ldots).  In \cite{Boyer:2010:spmv} we
developed some techniques to improve the \spmv operation in \linbox. Ideas
include the separation of the $\pm 1$ for removing multiplications, splitting
in a sum (HYB for hybrid format) of sparse matrix  whose formats are
independent and using specific routines. For instance, on $\modring{p}$ with
word size $p$, one can split the matrix ensuring no reduction is needed  in the
dot product and call Sparse \blas (from Intel \textsf{MKL} or Nvidia
\textsf{cuBLAS} for instance) on each matrix. One tradeoff is as usual between
available memory, time spent on optimizing \vs time spent on \apply, and all
the more so because we allow the concurrent storage of the transpose in an
optimized fashion, usually yielding huge speedups. This can be decided by
\adhoc optimizers.
% Helper structure to store a matrix as a sum of structures (HYB), possibly the
%
%\monitem{Parallelization}
\par
Work on parallelizations using \textsf{OpenCL}, \textsf{OpenMP} or
\textsf{XKaapi} for dense or sparse matrix multiplication include
\cite{Boyer:2010:spmv,WST12,DGPZ14}.
%
%%%%%%%%%%%%%%%%%%%%%%%%%%%%%%%
% Bench
%%%%%%%%%%%%%%%%%%%%%%%%%%%%%%%
% \input{Boyer_Brice/4-benchmarks}
\section{Benchmarking  for automated tuning and regression testing}\label{sec:bench}
Benchmarking was introduced in \linbox for several reasons. First, It
gives the user a convenient way to produce quality graphs with the
help
of a graphing library like \gnuplot
\footnote{\url{http://www.gnuplot.info/}}
and provides the \linbox website with automatically updated tables and graphs.
Second, it can be used for regression testing.  Finally, it will be used for
selecting default methods and setting thresholds in installation time autotuning. % A lot of libraries do some automatic tuning
%at installation (\textsf{fftw}, \textsf{ATLAS}, \textsf{NTL},\ldots).
%
\par
%
% What do we do differently ? Selection between "larger" algorithms, takes more time.
% Interpolation.
% XXX BTL\footnote{\url{http://projects.opencascade.org/btl/}}/eigen
%
%
\subsection{Performance evaluation and Automated regression testing}
Our plotting mechanism is based on two structures: {\tt PlotStyle} and {\tt
PlotData}. The  {\tt PlotGraph} structure uses the style and data to manage the
output.  We allow plotting in standard image formats, html and \LaTeX tables,
but also in raw csv or xml for file exchange, data comparisons and
extrapolation.
%JGD: no comprendo
This mechanism can also automatically create benchmarks in \linbox
feature matrix (this is a table that describes what solutions we support, on
which the fields).
% \danger dave benchmark formats discussion ?
%
\par
%
% \input{Boyer_Brice/fig_bench}
%
% XXX Time spent on each data is limited (will not start execution if fit
% (linear least squares) forecasts 'too long'.\\
%
% XXX Adapts to the environment.
%
Saving graphs in raw format can also enable automatic regression testing on the
buildbots that already checked our code. For some specifically determined matrices (of
various shapes and sizes and over several fields), we can accumulate the timings
for key solutions such as ({\tt rank}, {\tt det}, {\tt mul},\ldots) over time.
At each new release, when the documentation is updated, we can check any
regression on these base cases and automatically update the regression plots.
% \danger We need to implement this framework (not difficult; anybody?).
%JGD: we already have the buildbot, lets focus on this right now
%
\subsection{Automated tuning and method selection}
Some of the code in \linbox is already automatically tuned (such as thresholds
in \fgemm), but we improve on it.
%It is well known that CPU throttling while
%building ATLAS causes bad optimizations; the \fflasffpack library could also
%suffer from not very reliable threshold detection (up to $50\%$ relative
%difference for some thresholds between runs).
Instead of searching for a
threshold using fast dichotomous techniques, for instance, we propose to
interpolate curves and find the intersection. Using least squares fitting, we
may even tolerate outliers (but this is time consuming).
%JGD: there is already optimizer/winograd.C etc.
%
\par
Automatically tuning a library is not only about thresholds, it may
also involve method/algorithm selection. Our strategy is the following: a given
algorithm is tuned for each {\tt Helper} (method) it has.  Then the solution
(that uses these algorithms) is tuned for selecting the best methods.  At each
stage, defaults are given, but can be overridden by the optimizer. The areas
where a method is better are extrapolated from the benchmark curves.
%
%
%%%%%%%%%%%%%%%%%%%%%%%%%%%%%%%
% Conclusion
%%%%%%%%%%%%%%%%%%%%%%%%%%%%%%%
%
%
% \input{Boyer_Brice/5-conclusion}
%
%
%%%%%%%%%%%%%%%%%%%%%%%%%%%%%%%
% Bib
%%%%%%%%%%%%%%%%%%%%%%%%%%%%%%%
%
%
\fi % abstractonly
% \bibliographystyle{abbrv}
% \bibliographystyle{acm}
% unComment either of the following:
% \bibliography{bboyer}
% \input{Boyer_Brice/14_linbox_short.bbl}

%
%
% \end{document}
\end{document}